\documentclass[preprint2]{aastex}
\usepackage{graphicx}
\usepackage{color}

\begin{document}

\title{From the SpaceX Starlink megaconstellation to the search for Type-I civilizations}


\author{Osmanov Z.N.}
\affil{School of Physics, Free University of Tbilisi, 0183, Tbilisi,
Georgia}
\affil{E. Kharadze Georgian National Astrophysical Observatory, Abastumani, 0301, Georgia}

\begin{abstract}
Here we extrapolate the idea of launching SpaceX's Starlink satellites and study the possibility of building planetary megastructures (either designed as solid objects or as a web of satellites) by Type-I civilizations and the consequent detection of their techno-signatures. We have shown that the instruments of The Very Large Telescope Interferometer (VLTI) can potentially observe emission pattern of the huge constructions. Efficiency of the spectral variability method has been emphasized and the role of the FAST telescope was discussed.
\end{abstract}


\section{Introduction}

Last year on May $24$ the SpaceX launched the first set of $60$ Starlink satellites and up to now the total number is approximately $400$ aiming to reach at least $12000$ \citep{starlink} at the end of a decade program announced by Elon Musk. Such a huge number of satellites, distributed over almost the whole surface of the Earth might be considered as the first prototype of a possible megastructure around the Earth, which in principal, might be visible from the cosmos. Similarly, one may search for techno-signatures of alien civilizations.

The discovery of the Tabby's star \citep{kic846} and "Oumuamua" \citep{oum1} have provoked the revival of the search for extraterrestrial intelligence (SETI). The idea to search for techno-signatures of advanced alien societies have been proposed by  \cite{dyson}. Assuming that a civilization is advanced enough to build a megastructure around a host star to consume its whole energy, Dyson has concluded that such a huge (having the length-scale of the order of one AU) spherical construction - Dyson sphere (DS) - should be visible in the infrared (IR) spectrum. Civilizations harnessing the host star's total energy belong to the Type-II societies according to the classification by \cite{kardashev}. Type-I civilization is harnessing the total energy coming from the sun to the Earth. Our society is consuming less than the mentioned energy, therefore, an index, $K = log_{10}\left(P\right)/10-0.6$, introduced by \cite{shkl} for Earthlings is $0.7$, where $P$ denotes the average harnessed power in Watts. In the framework of the same classification Type-III is the alien high tech society which is able to use the total energy of the host galaxy. 

It is clear that detection of Type-II and Type-III civilizations is much easier than Type-I because of the much higher total consumed energies. Therefore, a special interest in the Dysonian SETI projects deserve Type-II,III techno-signatures and a series of papers are dedicated to identification of DS candidates \citep{timofeev,carrigan,gaia}. Dyson's original idea has been extended to hot DSs \citep{paper3,paper4} and the megastructures around pulsars \citep{paper1,paper2}. 

Despite high radiation intensity of Type-II,III technologies compared to Type-I techno-signatures, the latter still can be considered seriously in the SETI context. In particular, \cite{warming} studied effects of global warming as detectable biomarkers in Earth-like societies. 

Our civilization consumes approximately $1.5\times 10^{20}$ ergs s$^{-1}$ \citep{warming}, which is less than for Type-I society, $1.7\times 10^{24}$ ergs s$^{-1}$. If one assumes $1\%$ of an average growth rate of industry and the subsequent energy consumption, one can straightforwardly show that our civilization might reach Type-I in $\sim1000$yrs. It is quite probable that in $1000$yrs the level of technology will differ from ours, likewise ours is different from the one of middle ages. Therefore, one can assume that Type-I alien society is able to  cloak their planet by a sphere-like (or ring-like) structure to harness the total energy emitted from their host star toward the planet.

In this paper we consider the possible observational characteristics of a planetary megastructure partially or completely covering an Earth-like planet located in the habitable zone.

The paper is organized in the following way: in Sec. 2, we introduce main theory and study the techno-signatures of Type-I megaconstructions, obtaining major results and in Sec. 3 we outline them.

\section[]{Discussion and results}

At first we consider the question: energetically how feasible is launching material to cover an area comparable to the Earth's surface. As an example we examine the construction designed by means of Graphene as up to now the strongest material, our civilization is able to produce. Then, for the total mass of the megastructure we obtain
$$M\simeq 4\kappa\pi R_p^2\rho\Delta r\simeq$$
\begin{equation}
\label{mass} 
 \simeq 2\kappa\times 10^{17}\times\frac{\rho}{0.4g/cm^3}\times\frac{\Delta r}{1mm} \; g,
\end{equation}
where $\kappa\leq 1$ is a dimensionless geometric factor, $R_p$ is the Earth-like planet's radius ($R_{\oplus}\simeq 6400\;km$) density, $\rho$, is normalised by the Graphene density and it is assumed that since the Graphene is $10$ times stronger than steel the effective thickness of the construction, $\Delta r$, is normalised by $1$mm value. As it clear from here, the total mass is less than the total storage of Earth's Carbon, $1.85\times 10^{24}$ by seven orders of magnitude and thus one can find enough material to build a construction. Here we assume that the altitude is much less than $R_{\oplus}$. One can straightforwardly check that at geostationary orbit ($R\simeq 36000$km) the required mass becomes enormous compared to the above considered.

Another issue one should address, is to understand time scale required for launching such a huge mass of material. Roughly speaking, the civilization can start launching material from the times when its index was equal to ours, $0.7$ (See the introduction). Then, as we have already mentioned, to reach the Type-I level, it needs $1000$ yrs. The process will be feasible if the annual rate of growth of mass launching, $\mu$, is a small parameter. If one assumes that the first year's launched mass equals $M_0$, then in $1000$ yrs the total mass will be given by
\begin{equation}
\label{launch} 
M=M_0\left(1+\mu\right)^{1000}.
\end{equation}
After taking into account Eq. (\ref{mass}) and the fact that up to date there are $895$ satellites each with the mass $260$ kg \citep{starlink}, one can straightforwardly show that $\mu\simeq 0.021$. 

Generally speaking, the megastructure might be constructed by advanced Type-I civilization from extraterrestrial resources without launch costs \citep{haliki}, but let us consider the worst case: the Solar energy is stored and then utilized to launch material on the orbit.

In this case energy required to launch material on the altitude $H\sim 500$km is as follows
\begin{equation}
\label{energy} 
E\simeq \frac{GMM_pH}{R_{\oplus}^2}\simeq 5.7\kappa\times 10^{27}\; ergs,
\end{equation}
where $M_p = M_E $ is Earth-like planet's mass and $M_E\simeq 6\times 10^{27}$ g is the Earth's mass. 

On the other hand, power from the host star toward the planet is given by
\begin{equation}
\label{power} 
P\simeq \frac{L}{4}\times\left(\frac{R_p}{r}\right)^2\simeq 1.7\kappa\times 10^{24}\; ergs\;s^{-1},
\end{equation}
where $L\simeq 3.8\times 10^{33}$ergs s$^{-1}$ is the solar-type star's luminosity and $r\simeq 1.5\times 10^{13}$ cm is the radius of the habitable zone. Whatever the propulsion mechanism (electromagnetic launch or some other mechanisms), it is evident from Eq. (\ref{power}) that the energy required to launch a thin shell around a planet can be extracted in approximately ten hours with $10\%$ of efficiency of energy conversion. If the megastructure is a web of many satellites orbiting the planet, then the kinetic energy should be added to the aforementioned value, leading to ten times more total energy, which, energetically is quite feasible for Type-I alien societies.

For a uniformly distributed spherical mass such a construction will not require any energy to stabilise it. Unlike this scenario, if instead of the sphere one uses a ring, the planet will be characterised by the out of plane stability, whereas for in-plane displacements the dynamics is unstable \citep{paper1}. In this work we have discussed that such rings may have sense only if the power required to maintain stability is small compared to the total received power, leading to the following condition for maximum displacement from equilibrium
\begin{equation}
\label{ksi} 
\xi<<\frac{0.37}{R_p}\left(\frac{P}{M}\right)^{1/2}\left(\frac{2R_p^3}{GM_p}\right)^{3/4}\simeq 0.07,
\end{equation}
where $\xi\equiv d/R_p$ and $d$ denotes the displacement of the ring from the equilibrium position. We would like to emphasise that such a precision is not difficult to achieve: in the Lunar laser ranging experiment\footnote{Data is avalable from the Paris Observatory Lunar Analysis Center: http://polac.obspm.fr/llrdatae.htm}
the distance is measured with the precision of the order of $10^{-10}$. 

By assuming that the whole emission incident on the megastructure is absorbed, one can easily show that the temperature is given by
\begin{equation}
\label{temp} 
T\simeq\left(\frac{P}{16\sigma r^2}\right)^{1/4}\simeq 280 K,
\end{equation}
leading by means of the Wien's law to the emission peak in the IR spectral band with the following wavelength
\begin{equation}
\label{lambda} 
\lambda = \frac{b}{T}\simeq 10 \mu m,
\end{equation}
where $\sigma$ denotes the Stefan-Boltzmann constant and $b\simeq 2898\mu$mK.
The average value of the flux can be estimated as follows
\begin{equation}
\label{flux} 
F\simeq\frac{P}{4\pi D^2}\simeq 1.6\times10^{-17}\times\left(\frac{100 \;ly}{D}\right)^2 ergs\;s^{-1} cm^{-2},
\end{equation}
where the distance to the extrasolar system, $D$, is normalised by $100$ light years. The Very Large Telescope Interferometer (VLTI) could detect such fluxes. In particular, the limiting flux value of the VLTI instruments\footnote{https://www.eso.org/sci/facilities/paranal/telescopes/vlti} in $1$ hour is of the order of $2.3\times 10^{-18} ergs\;s^{-1} cm^{-2}$. Therefore, potentially one can monitor the spherical volume with radius of the order of $R_0  = 260\; ly$. In the Solar neighbourhood the stellar number density, $n$, is of the order of $0.138$ star pc$^{-3}$ \citep{dens}, which means that the total number of stars could be
\begin{equation}
\label{numb} 
N\simeq\frac{4\pi R_0^3}{3}\; n\simeq 3\times 10^5,
\end{equation}
with approximately $1000$ G-type (Solar-type) stars. Although the flux method cannot distinguish emission from the planet and the megastructure, it allows to find objects in the habitable zone. For identifying the candidates of megaconstruction one can use the spatial resolution of the telescopes.

The VLTI telescope has the maximum angular resolution, $\theta_m$, of the order of $0.001$ mas (milliarcsecond - please see the technical characteristics of the VLTI instruments. By taking into account this value, one can obtain the maximum resolving distance 
\begin{equation}
\label{dist} 
D_m = \frac{2R}{\theta_m}\simeq 280\;ly.
\end{equation}
From the aforementioned two critical values, it is clear that planetary-scale megastructures inside $260\; ly$ might be detectable. 

It is worth noting that a monolithic DS will be less stable compared to concentric rotating rings. In this case very interesting observational features might arise. By means of the non-relativistic Doppler effect, the observed wavelength is given by \citep{carroll}
\begin{equation}
\label{dopp1} 
\lambda \simeq \lambda_s\left(1+v\cos\theta/c\right),
\end{equation}
where by $\lambda_s$ we represent the original wavelength emitted by the ring, $\theta$ is an angle of velocity direction measured in an observer's frame of reference, $v$ denotes the corresponding velocity and $c$ is the sped of light. Without going into geometric details, to make the orders of magnitude, we consider diametrically opposite sides of the ring relative to an observer moving with velocities $\upsilon_p \pm v_{_I}$, where $\upsilon_p$ denotes the orbital velocity of the planet around a host star and $v_{_I}=\sqrt{GM_p/R_p}$ is the orbital speed of the ring around a planet. Then, for the diametrically opposite sides of the ring from Eq. (\ref{dopp1}) one obtains
\begin{equation}
\label{dopp2} 
\frac{\lambda_s}{\Delta\lambda}\simeq\frac{c}{2\upsilon_{_I}}\simeq 1.9\times 10^4\times \left(
\frac{R_p}{R_{\oplus}}\times \frac{M_{\oplus}}{M_p} \right)^{1/2}.
\end{equation}
In this regard, one should note that up to now the VLTI has the highest resolving power $RP\equiv\lambda/\Delta\lambda = 25000$ for $\lambda \approx 10 \mu$m (which corresponds to the blackbody radiation with $T=300$K) \footnote{www.eso.org/public/teles-instr/paranal-observatory/vlt/} and since $RP>\lambda_s/\Delta\lambda$, such orbital motions of the ring might be detected by the spectrographs of the VLTI instruments. Similarly, if instead of a solid mega-construction one uses a web of satellites, the same method might be used. It is worth noting that the observational pattern might be characterized by absorption lines of the matter the megastructure is made of. Therefore, in principle, the emission fingerprint might carry information about the matter, but this is not the scope of the current paper and we do not consider it.

One has to note that the SpaceX starlink satellites use radio communication. Therefore, a reasonable question might appear: is it possible to detect their radio signals by means of the China's FAST telescope? By taking into account that its system temperature is $T_{sys}\simeq 20$K and the illuminated aperture area, $A\simeq 7108$ cm$^2$ \citep{fast}, one can conclude that the minimum spectral flux density, $kT_{sys}/A$, which can be distinguished from noise is of the order of $6\times 10^{-24}$ ergs s$^{-1}$cm$^{-2}$Hz$^{-1}$. On the other hand, by assuming that only $1\%$ of the incident Solar energy is used for interstellar communication, one can find that if one uses the Hydrogen atom's frequency, 1420 Hz, the FAST telescope can detect such isotropic sources up to the distances of the order of $160$ pc.

Making use of simple geometry we have performed our calculations for solar-type stars with Earth-like planets and it is clear that the similar estimates can be straightforwardly performed for Type-I civilizations living in the systems with different parameters. 

\section{Conclusion}
Extrapolating the idea of SpaceX's Starlink constellation we assume that an alien society with index $K = 0.7$ (our civilization) will reach Type-I in $1000$ years, which is enough to build a planetary megastructure for collecting the required material.

Launching has been analysed from the point of view energy costs and it has been shown that energy required to construct a megastructure is small compared to the total energy received from the star.

We have shown that the construction will be visible in the IR spectrum, which might be detected by VLTI instruments up to the distance $260$ light years, with $\sim 10^3$ Solar-type stars.

We have also emphasised that the spectral variability method might be an efficient tool to detect either orbital rotation of the solid megastructure or internal motions of small satellites.

It has been estimated the possibility to detect the radio emission operating on the Hydrogen atom?s frequency. It has been found that for reasonable parameters FAST can detect such radio sources from relatively large distances.

\section*{Acknowledgments}
The research was supported by the Shota Rustaveli National Science Foundation grant (NFR17-587). 

\end{document}